\documentclass[epjST]{svjour}
\usepackage{graphics}

\newcommand{\apjl}{Astrophys. J. Lett.}%
\newcommand{\mnras}{Mon. Not. Roy. Astron. Soc.}%
\newcommand{\lrr}{Living Reviews in Relativity}%

\newcommand{\prl}{Phys. Rev. Lett.}    
\newcommand{\prd}{Phys. Rev. D}    
\newcommand{\prc}{Phys. Rev. C}    
\newcommand{\apj}{Astrophys. J.}

\begin{document}

\title{Impact of quark deconfinement in neutron star mergers and hybrid star mergers}

\author{Andreas Bauswein\inst{1}\fnmsep\thanks{\email{a.bauswein@gsi.de}} \and Sebastian Blacker\inst{1,2}\fnmsep\thanks{\email{s.blacker@gsi.de}}}
\institute{GSI Helmholtzzentrum f\"ur Schwerionenforschung, Planckstra{\ss}e 1, 64291 Darmstadt, Germany \and Technische Universit\"at Darmstadt, Fachbereich Physik, Institut f\"ur Kernphysik, Schlossgartenstraße 9, 64289 Darmstadt, Germany }
\abstract{We describe an unambiguous gravitational-wave signature to identify the occurrence of a strong phase transition from hadronic matter to deconfined quark matter in neutron star mergers. Such a phase transition leads to a strong softening of the equation of state and hence to more compact merger remnants compared to purely hadronic models. If a phase transition takes place during merging, this results in a characteristic increase of the dominant postmerger gravitational-wave frequency relative to the tidal deformability characterizing the inspiral phase. By comparing results from different purely hadronic and hybrid models we show that a strong phase transition can be identified from a single, simultaneous measurement of pre- and postmerger gravitational waves. Furthermore, we present new results for hybrid star mergers, which contain quark matter already during the inspiral stage. Also for these systems we find that the postmerger GW frequency is increased compared to purely hadronic models. We thus conclude that also hybrid star mergers with an onset of the hadron-quark phase transition at relatively low densities may lead to the very same characteristic signature of quark deconfinement in the postmerger GW signal as systems undergoing the phase transition during merging.}

\maketitle
\section{Introduction}
\label{intro}

Neutron stars are formed by the material of the inner core of massive stars, which undergo a gravitational collapse. The enormous gravitational attraction compresses this matter to very high densities such that neutron stars contain more than a solar mass within a diameter of about 25~km. In fact, neutron stars are the places in the universe, where the highest densities can be found in stable equilibrium. Comparable densities can only be obtained in heavy-ion collisions for a very short period of time, e.g.~\cite{Friman2011}.

For this reason neutron stars are very promising objects to search for deconfined quark matter, which for low temperatures is expected to occur at some density beyond nuclear saturation. Whether or not neutron stars feature a quark core, is currently unclear. It is also not known how strongly the transition from ordinary baryonic matter to deconfined quark matter affects the stellar structure. If the phase transition is connected with a considerable density jump (latent heat), the mass-radius relation of compact stars may feature a pronounced kink as consequence of the sudden softening of the equation of state (EoS). If the transition to deconfined quark matter proceeds in a more continuous way, the impact on the stellar structure may be rather minor and in fact hard to detect \cite{Alford2005}. 

Gravitational waves (GWs) provide a very promising tool to detect traces of quark matter if it exists in neutron stars (e.g.~\cite{Orsaria2019}), in which case these compact objects are often called hybrid stars. In particular, the merging of neutron stars and the associated GW signal may reveal the presence of quark matter because the collision is a highly dynamical event, which is strongly influenced by the stellar structure and the EoS of high-density matter.

In this contribution we discuss an unambiguous signature of a phase transition to deconfined quark matter in the GW signal of neutron star mergers~\cite{Bauswein2019}. We emphasize the importance of providing evidence that a specific signature is generated only by the presence of quark matter. To this end it is critical to show that all viable hadronic models do not produce such characteristics. We also remark discuss how the presence or absence of such a signature of deconfined quark matter can constrain the onset density of the quark-hadron phase transition~\cite{Bauswein2019,Blacker2020}. 

As a completely new finding, we discuss the merging of two hybrid stars, i.e. systems which contain quark matter already before merging. We find that such hybrid mergers produce a signature similar to that of systems which undergo a phase transition to quark matter only during the collision. This is in particular important in a scenario where the onset of quark deconfinement occurs already at very low densities, e.g.~\cite{Blaschke2020}.

In Sect.~\ref{dynamics} we present the dynamics of neutron star mergers and highlight how the EoSs determines the outcome of the merger and the GW signal produced before and after the merging.
Sect.~\ref{Postmerger} discusses the influence of a strong first-order phase transition on the postmerger GW signal. In subsection \ref{identify} we focus on how the GW signal can reveal the occurrence of a strong first-order phase transition in the merger remnant. In this context we highlight the outcome and GW signature of hybrid mergers (subsection~\ref{hybrid}). We summarize and conclude in Sect.~\ref{Summ}.

\section{Dynamics of neutron star mergers}\label{dynamics}
There are different features of the GW signal which contain information about the properties of high-density matter. The phase before the merging of the binary components is called inspiral referring to the trajectories of the stars, which orbit around each other while the orbital separation shrinks. The GW signal of the inspiral is dominantly shaped by the orbital motion \cite{Faber2012}. The stellar structure affects the orbital dynamics such that less compact stars lead to an accelerated inspiral compared to a system of point particles of the same mass. This effect is described by the tidal deformability $\Lambda=\frac{2}{3}k_2\left(\frac{c^2 R}{G M}\right)^5$ with the stellar mass $M$ and radius $R$. $k_2$ is the tidal Love number \cite{Hinderer2008} ($c$ and $G$ are the speed of light and the gravitational constant). The tidal deformability is the EoS parameter that can be extracted from the GW signal \cite{Abbott2017}, see~\cite{Chatziioannou2020} for a review.
\begin{figure}
\begin{center}
\resizebox{0.75\columnwidth}{!}{%
 \includegraphics{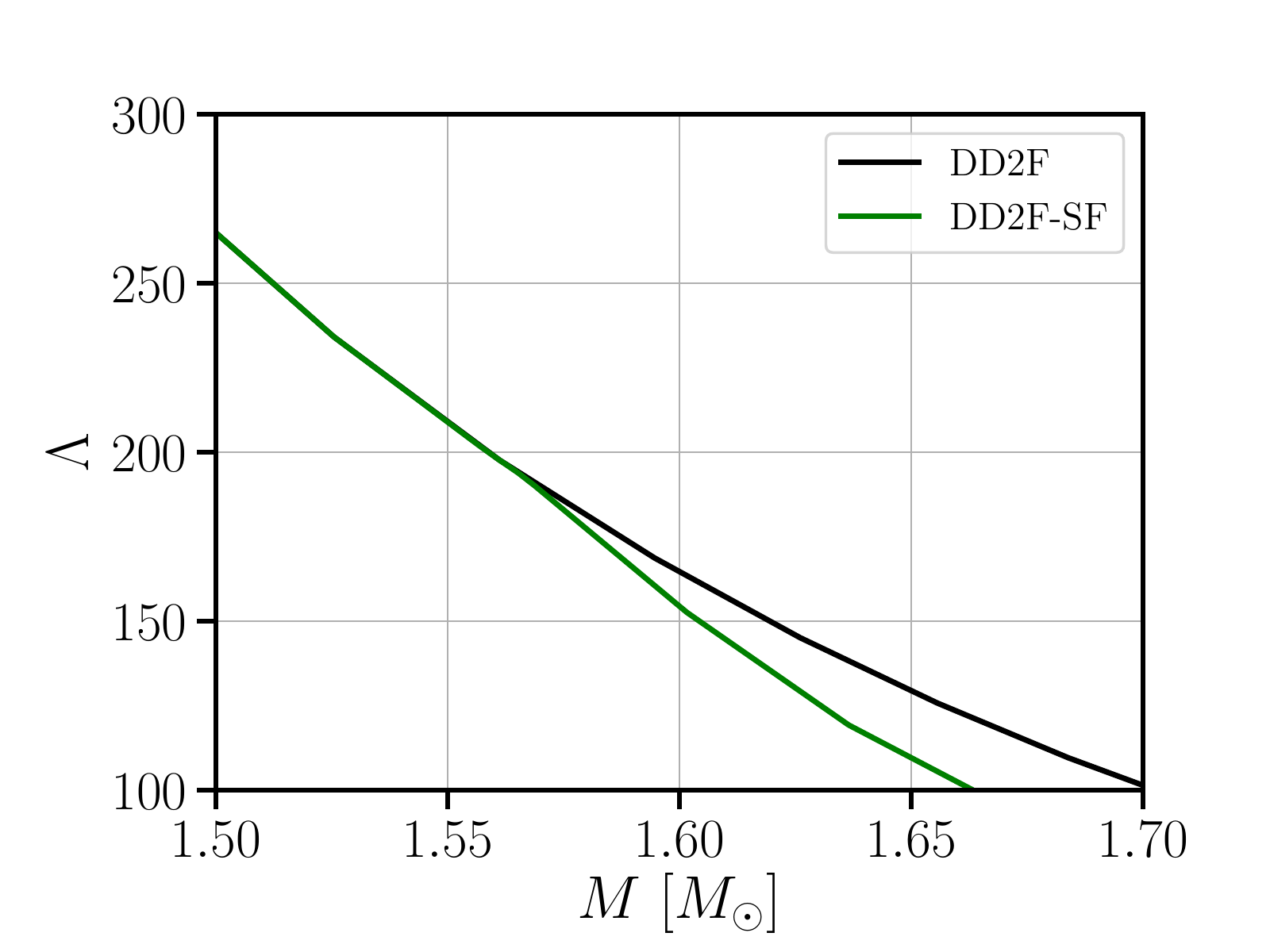} }
\caption{Tidal deformability $\Lambda$ of an isolated neutron star as a function of mass $M$ for the purely hadronic DD2F EoS (black curve) and the hybrid DD2F-SF-1 EoS (green curve).}
\label{fig:lambda}       
\end{center}
\end{figure}

As apparent from its definition, $\Lambda$ scales tightly with the stellar radius. The occurrence of quark matter beyond some threshold mass leads to a kink in the $\Lambda-M$ relation similar as in the mass-radius relation. An example is provided in Fig.~\ref{fig:lambda}, which shows $\Lambda(M)$ for a purely hadronic EoS (black curve) and for a hybrid model (green curve). The density regime below the onset density of the phase transition is described by the same hadronic model. Hence, the two curves coincide for masses below the threshold mass which marks the occurrence of deconfinement. For these models quark matter appears at a mass of about 1.57~$M_\odot$, where a characteristic kink is visible.

The hadronic regime of the models shown here is described by the DD2F EoS~\cite{Typel2010,Alvarez-Castillo2016} and the hybrid EoS is based on the string-flip model of \cite{Fischer2018,Bastian2018}. Additional details on these two EoS models and their parameters can be found in the supplemental material of \cite{Bauswein2019} and the references therein.

Detecting such a kink in $\Lambda(M)$ would indicate the presence of a phase transition. However, Fig.~\ref{fig:lambda} shows that it may be very difficult to identify a phase transition based entirely on the GW inspiral signal. It would require the detection of different binary mergers with slightly different masses above and blow the kink. Finite-size effects on the inspiral GW signal become smaller at higher masses because the binary components are more compact and thus behave more similar to point particles. Thus, the extraction of the tidal deformability for high-mass mergers is challenging. Moreover, measurements of $\Lambda$ will contain a systematic and a statistical error, both of which may be too large to actually identify the kink, which would require a measurement uncertainty of at most 5\% (see, however, \cite{Chen2019,Chatziioannou2019} for ideas to infer the presence of a phase transition from combining many detections).

Apart from the inspiral phase, the EoS also affects the outcome of a binary merger. The collision of the binary components forms a massive, rotating remnant, whose stability is determined by the properties of high-density matter \cite{Bauswein2017a}. The merged object may be (temporarily) stable even if its total mass exceeds the maximum mass of non-rotating neutron stars because of the strong centrifugal support by the rapid rotation. In fact, for most EoS models the formation of a rotating neutron star remnant is expected within the mass range of typical neutron star binaries around 2.7~$M_\odot$. However, for total binary masses beyond some threshold binary mass the remnant will undergo a prompt collapse to a black hole~\cite{Bauswein2013,Bauswein2020}.

\begin{figure}
\begin{center}
\resizebox{0.75\columnwidth}{!}{%
 \includegraphics{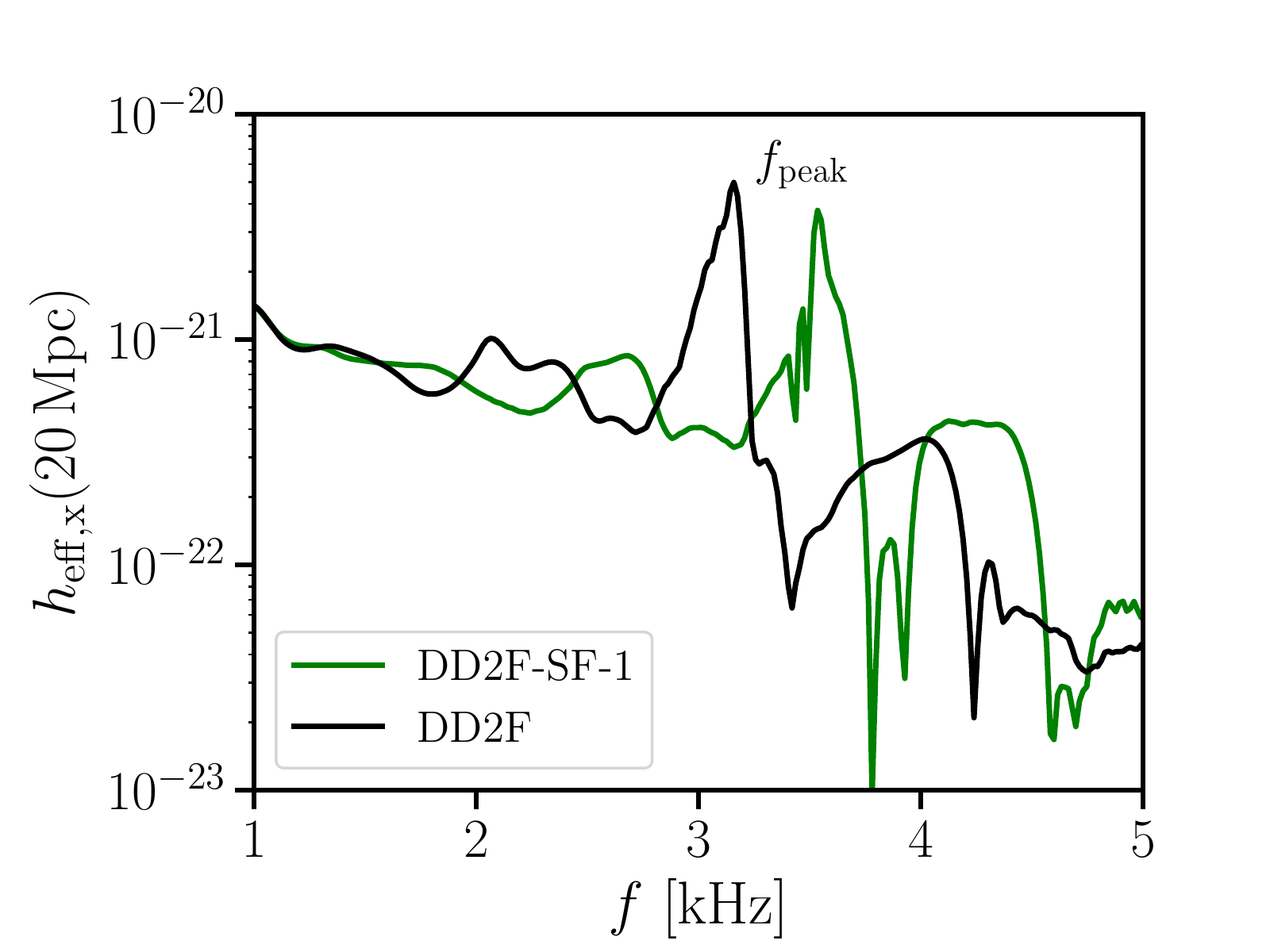} }
\caption{Gravitational wave spectrum of the cross polarization at a distance of 20~Mpc along the polar axis for the purely hadronic DD2F model (black curve) and the hybrid DD2F-SF-1 EoS (green curve). Figure taken from \cite{Bauswein2019}.}
\label{fig:spectrum}       
\end{center}
\end{figure}

For systems without direct black hole formation the GW emission of the rotating neutron star remnant provides another opportunity to learn about the properties of high-density matter \cite{Bauswein2012}. The collision itself leads to strong oscillations of the merger remnant, which produce GWs in the kHz range. An example is given in Fig.~\ref{fig:spectrum}, which shows the GW spectrum of a 1.35-1.35~$M_\odot$ merger for the two different EoS models used for Fig.~\ref{fig:lambda}. One can recognize different features in the spectrum, i.e. different frequency peaks, which correspond to different oscillation modes of the remnant and are very characteristic of the EoS \cite{Bauswein2019a}. As such signals are a result of a dynamical evolution, it requires detailed relativistic hydrodynamical simulations in three dimensions to follow the merging process and to compute the corresponding GW emission (see e.g.~\cite{Bauswein2019b,Blacker2020} for snapshots from simulations).

These simulations also reveal that the merging results in a strong density increase. See e.g. \cite{Bauswein2019,Blacker2020} for a figure showing the time evolution of the maximum density. The density increase also implies that the GW signal of the postmerger stage carries information from the high-density regime of the EoS. While the inspiral signal only contains information about the EoS regime of the progenitor stars, the consideration of the postmerger GW signal is a natural choice to search for the impact of a phase transition to deconfined quark matter which may occur at higher densities \cite{Most2019,Bauswein2019,Weih2020,Bauswein2020,Blacker2020}. Generally, the impact of the EoS on the postmerger phase can be understood as follows. The EoS determines the stellar structure of the remnant, which affects the frequencies of the different oscillation modes. EoSs which are soft and thus result in compact stars and compact merger remnants, will produce GW emission at higher frequencies. Indeed, it is known that the dominant oscillation frequency scales with the size of the merger remnant and with the size of non-rotating neutron stars \cite{Bauswein2012a}.

\section{Postmerger gravitational wave emission and first-order phase transitions}\label{Postmerger}
Because the GW signal of a merger probes two different density regimes (before and after the merger), it can reveal a strong phase transition occurring in the merger remnant \cite{Bauswein2019}.

The spectra in Fig.~\ref{fig:spectrum} show the postmerger GW emission from a merger described by a hadronic EoS (black) and from an event where a phase transition takes place in the merger remnant (green). Note that both EoSs are identical at densities below the phase transition. One can clearly see that the hybrid EoS with a phase transition to deconfined quark matter leads to higher postmerger frequencies. This is understandable because the softening of the EoS leads to a more compact remnant which enhances the postmerger GW frequencies.

The dominant postmerger oscillation frequency $f_\mathrm{peak}$ of the hybrid model is shifted by about 450~Hz with respect to the hadronic model. Note that $f_\mathrm{peak}$ is caused by the fundamental quadrupole fluid mode and is a robust feature of neutron star merger simulations~\cite{Stergioulas2011}. Typically, it lies between 2~kHz and 4~kHz, depending on the total mass of the system and the EoS \cite{Bauswein2012,Bauswein2012a,Hotokezaka2013,Takami2014,Bernuzzi2015,Bauswein2015}. Since $f_\mathrm{peak}$ is expected to be measurable in the near future with enhanced GW detectors \cite{Clark2016,Chatziioannou2017,Torres-Rivas2018,Easter2019}, it is a quantity that can indicate the presence a of strong phase transition in merger remnants.

\begin{figure}
\begin{center}
\resizebox{0.75\columnwidth}{!}{%
 \includegraphics{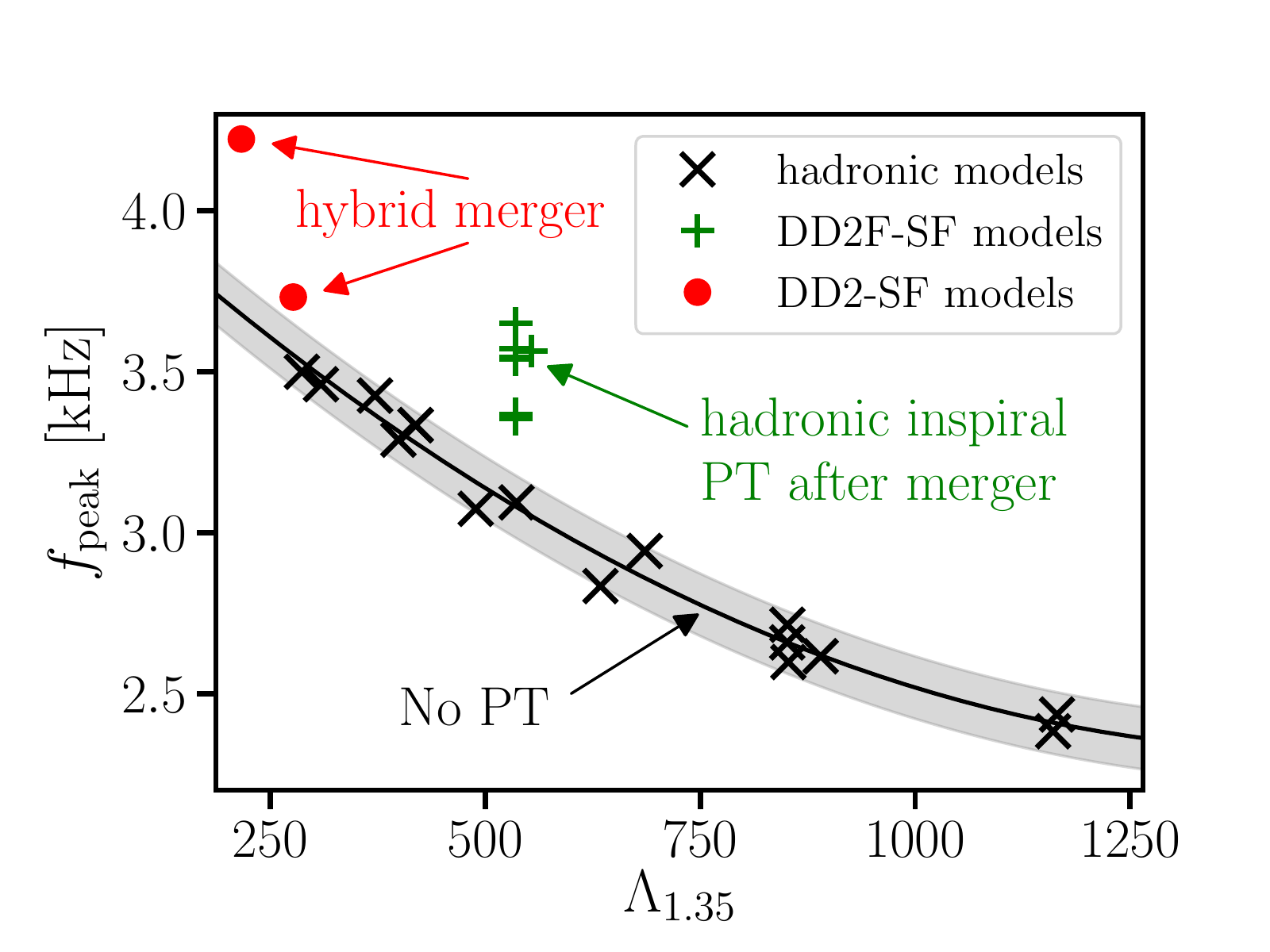} }
\caption{Dominant postmerger GW frequency $f_\mathrm{peak}$ as a function of the tidal deformability $\Lambda$ for 1.35--1.35~$M_{\odot}$ mergers with different microphysical EoSs. Black crosses refer to results with purely hadronic EoSs while green plus signs depict results from hybrid DD2F-SF models. For these hybrid models the phase transition takes place during the merger, hence the inspiraling stars are still purely hadronic. Red circles display results from hybrid DD2-SF models. These hybrid EoSs have very low onset densities such that the inspiraling neutron stars are hybrid stars (see Fig.~\ref{fig:MR} for the mass-radius relations). The solid curve is a least squares fit with a second-order polynomial to the data excluding results from hybrid EoSs. The gray shaded area illustrates the largest deviation of the data of purely hadronic models from the fit. Results falling within this area are consistent with the assumption of a purely hadronic EoS, while outliers are interpreted as evidence for a strong phase transition. The results from hadronic and DD2F-SF models are taken from \cite{Blacker2020}.}
\label{fig:flam}       
\end{center}
\end{figure}
However, a large value of $f_\mathrm{peak}$ by itself may not necessarily be indicative of a phase transition. For instance, for this binary mass configuration $f_\mathrm{peak}$ around and above 3.5~kHz can also be found for very soft hadronic EoSs~\cite{Bauswein2012}. 

\subsection{Identifying a strong first-order phase transition}\label{identify}
An unambiguous signature of a strong phase transition can be obtained by comparing the tidal deformability $\Lambda$ with $f_\mathrm{peak}$~\cite{Bauswein2019}. Such a comparison is shown in Fig.~\ref{fig:flam}. It compiles $f_\mathrm{peak}$ as a function of $\Lambda$ for simulations of 1.35--1.35~$M_\odot$ mergers with many different EoSs. See \cite{Blacker2020} for a list of hadronic and hybrid EoS models used in this figure; two additional hybrid models with an early onset of quark deconfinement are included here for the first time and further discussed below. All EoS used for Fig.~\ref{fig:flam} are compatible with current observational constraints by GW170817 and by the maximum-mass limit set by pulsar observations \cite{Abbott2017,Bauswein2017,Antoniadis2013,Cromartie2019}.

In Fig.~\ref{fig:flam}, $\Lambda$ refers to the tidal deformability of the individual inspiraling star. Note that for equal-mass binaries $\Lambda$ coincides with the effective tidal deformability $\tilde{\Lambda}=\frac{16}{13}\frac{(M_1+12 M_2)M_1^4\Lambda_{1}+(M_2+12 M_1)M_2^{4}\Lambda_{2}}{(M_1+M_2)^{5}}$ of the binary system, which is the parameter that is actually measure by GW observations (see~\cite{Chatziioannou2020}). Here, $M_i$ and $\Lambda_i$ refer to the mass and the tidal deformability of the individual stars, respectively. For the sake of clarity we omit a distinction in the following and refer to~\cite{Blacker2020} for further discussions. Black crosses show results from simulations with purely hadronic EoS models. The solid black line is a least squares fit to data from those hadronic models with a second-order polynomial. The gray shaded area visualizes the maximum deviation of this data from the fit (the fit parameters can be found in \cite{Blacker2020}). It is apparent that for these models $f_\mathrm{peak}$ scales tightly with $\Lambda$ and that this relation is well described by the displayed fit. Note that three of the hadronic EoSs contain a phase transition to hyperonic matter. However, they still follow the $f_\mathrm{peak}$--$\Lambda$ relation which indicates that this transition does not strongly influence the remnants structure. See \cite{Blacker2020} for more details.

The green plus signs display simulation results with different hybrid EoSs. One can see that the postmerger frequencies clearly deviate from the tight $f_\mathrm{peak}$--$\Lambda$ relation valid for purely hadronic EoSs. The postmerger frequencies of hybrid models are significantly shifted towards higher frequencies. In these simulations the phase transition occurs during merging. This implies that $f_\mathrm{peak}$ is affected by the occurrence of deconfined quark matter while $\Lambda$ is not. Note that all of these hybrid models are based on the same underlying hadronic reference EoS model (at densities below the onset density of quark deconfinement), which is why they all have the same value of the tidal deformability $\Lambda$.

If $f_\mathrm{peak}$ exceeds the empirical $f_\mathrm{peak}$--$\Lambda$ relation by more than 200~Hz (more than the maximum deviation of a hadronic model from the fit in Fig.~\ref{fig:flam}), this is a clear, unambiguous signature of a strong first-order phase transition, since all hadronic models behave differently. Also note that, as stated before, a large $f_\mathrm{peak}$ value by itself is not indicative of a phase transition since other hadronic models can lead to $f_\mathrm{peak}$ values comparable to those from hybrid EoSs. Only the comparison of $f_\mathrm{peak}$ and $\Lambda$ with the empirical  $f_\mathrm{peak}$--$\Lambda$ relation can reveal the phase transition.

Importantly, also for other binary mass configurations including asymmetric mergers such a signature can be observed. See~\cite{Blacker2020} for similar relations with other total binary masses and mass ratios different from unity.

\subsection{Hybrid star mergers}\label{hybrid}

Figure~\ref{fig:flam} also shows simulation results from two other hybrid EoSs (red circles). These hybrid models are based on a different underlying hadronic model DD2 \cite{Typel2010,Hempel2010}, but the same quark model as for the DD2F-SF EoSs is employed~\cite{Bastian2018,Fischer2018}. In particular, these EoSs feature a very early onset of quark deconfinement such that the phase transition to deconfined quark matter occurs already in relatively light stars below 1~$M_\odot$. 

\begin{figure}
\begin{center}
\resizebox{0.75\columnwidth}{!}{%
 \includegraphics{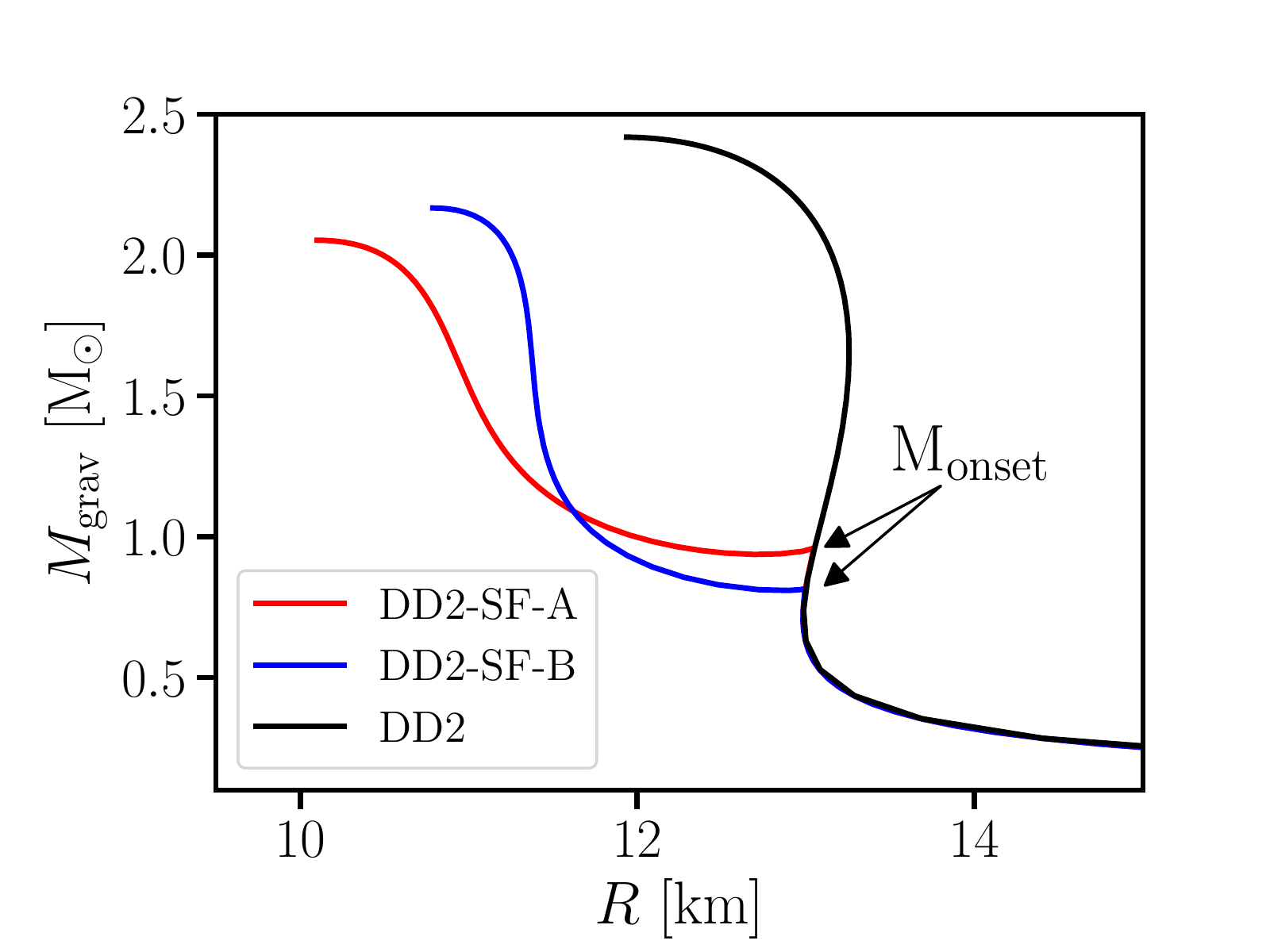} }
\caption{Mass-radius relations of cold, non-rotating neutron stars with hybrid DD2-SF-A/B models (colored curves) together with the purely hadronic model DD2. A phase transition to deconfined quark matter leads to a kink in the relations and typically to more compact neutron stars. One can see that $M_\mathrm{onset}$, i.e. the minimal neutron star mass with deconfined quark matter in the core is below 1~M$_\odot$ for both models. Below $M_\mathrm{onset}$ the two curves coincide with the DD2 mass-radius relation.}
\label{fig:MR}       
\end{center}
\end{figure}

Fig.~\ref{fig:MR} shows the mass-radius relations of these two models together with the underlying hadronic model DD2. As one can see, the hybrid EoSs have relatively low onset densities of the phase transition. The lightest resulting neutron stars with deconfined quark matter in their cores have masses of M$_\mathrm{onset}$ = 0.96~$M_{\odot}$ and M$_\mathrm{onset}$ = 0.82~$M_{\odot}$, respectively.

There is likely a lower mass limit of neutron stars because during their formation a certain threshold mass should be exceeded to trigger a gravitational collapse. Hence, for such an EoS all neutron stars are expected to be in fact hybrid stars containing a quark matter core. This obviously implies that also binary mergers are in fact hybrid star mergers. As apparent from Fig.~\ref{fig:MR} and the previous discussion, in this case also the inspiral GW signal is affected by the presence of quark matter. Specifically, this implies that the tidal deformability carries information about the phase transition. It thus arises the question whether or not a comparison between $\Lambda$ and $f_\mathrm{peak}$ does indicate the presence of quark deconfinement within such a scenario. Recall that for the other hybrid models discussed above, a sudden softening of the EoS by the phase transition during merging is responsible for the increase of the postmerger frequency. This effect may not be expected if quark matter is already present before merging.

However, as one can see from Fig.~\ref{fig:flam} the red circles still deviate from the empirical $f_\mathrm{peak}$--$\Lambda$ relation for purely hadronic EoSs by more than 200~Hz. This means that the identification of a strong phase transition is still possible even for the merger of two hybrid stars. Since temperatures rise during merging, we speculate that here also temperature effects play a crucial role in increasing the postmerger frequency. On the one hand, thermal pressure support in quark matter may be reduced in the sense that an effective thermal ideal-gas index is smaller than for ordinary nuclear matter, which leads to an increase of $f_\mathrm{peak}$ (see~\cite{Bauswein2010}). On the other hand, finite temperatures may shift the phase boundaries to lower densities implying that the quark matter core grows significantly and leads to an additional compactification of the remnant. As these simulations represent only a very first explorative study of this scenario, future work should consider an even larger set of hybrid EoSs of this type to understand their detailed impact on the GW signal. 

\begin{figure}
\begin{center}
\resizebox{0.75\columnwidth}{!}{%
 \includegraphics{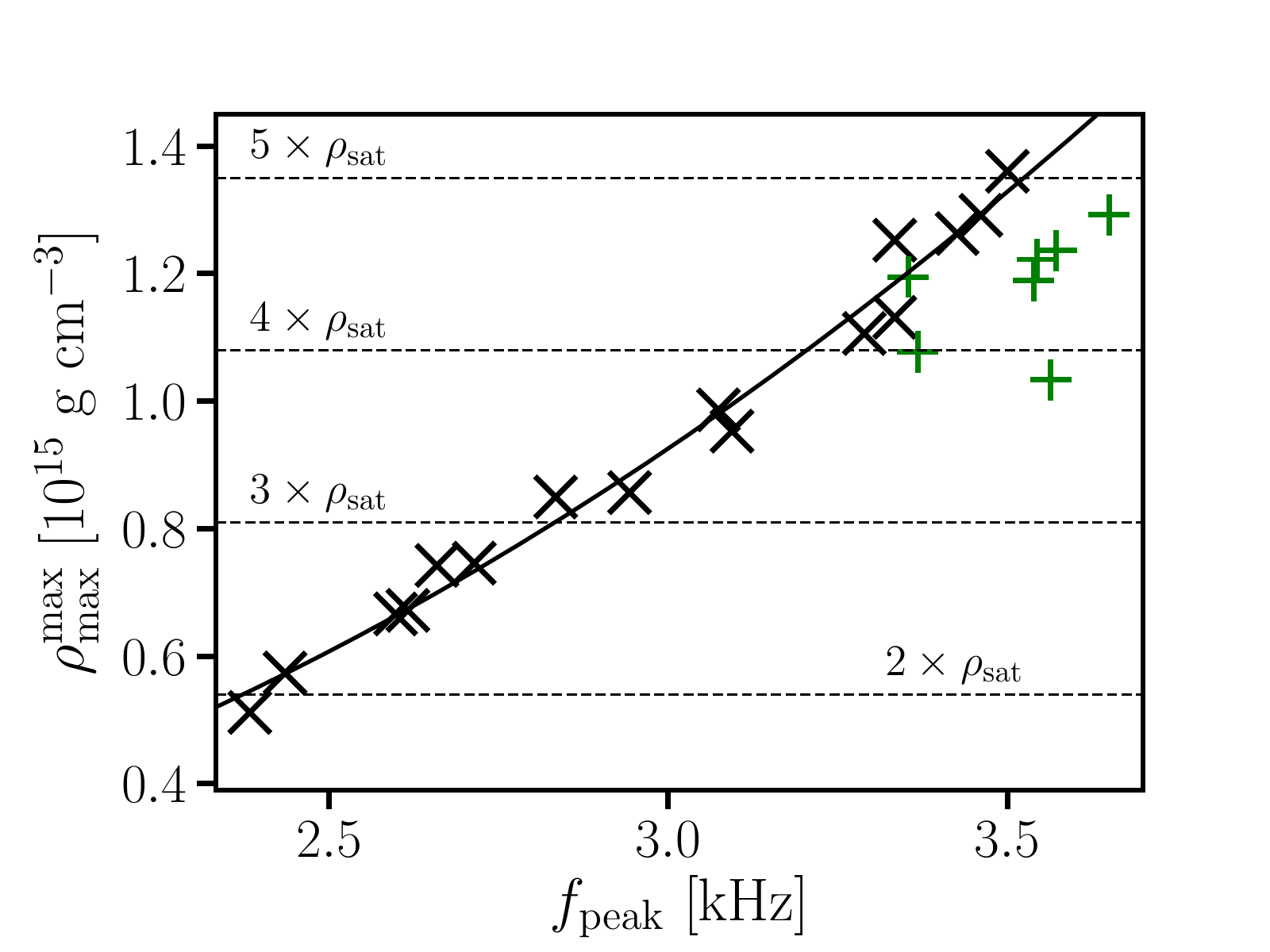} }
\caption{Maximum rest-mass density $\rho^\mathrm{max}_\mathrm{max}$ during the first 5 milliseconds after merging as a function of the dominant postmerger GW frequency $f_\mathrm{peak}$ for 1.35--1.35~$M_{\odot}$ mergers with different microphysical EoSs. Black crosses refer to results with purely hadronic EoSs while green plus signs depict results from hybrid DD2F-SF models. The solid curve is a least squares fit to the data excluding results from hybrid DD2F-SF models. Figure taken from \cite{Blacker2020}.}
\label{fig:frho}       
\end{center}
\end{figure}

Before concluding we emphasize that the relations discussed here are also useful to place a constraint on the onset density of quark deconfinement. This is based on the observation that measurable GW features like $f_\mathrm{peak}$ also inform about the densities which are reached in the postmerger remnant (see Fig.~\ref{fig:frho}). Hence, in a first step the aforementioned signature can be employed to determine whether or not quark deconfinement took place in a merger. Then, the relations between the maximum density that was reached in the merger remnant and the dominant oscillation frequency can be employed to place upper or lower bound on the onset density of quark deconfinement. We refer to~\cite{Blacker2020} for a concrete outline of this procedure and an extensive discussion of the involved subtleties.

\section{Summary and discussion}\label{Summ}
We have discussed an unambiguous and measurable GW signature to identify a strong phase transition to deconfined quark matter occurring in neutron star mergers. We show that a comparison between the tidal deformability and the dominant postmerger frequency can provide strong evidence for the presence of a phase transition. The tidal deformability $\Lambda$ characterizes the impact of the EoS during the pre-merger inpsiral phase and can be infered from the corresponding GW signal. The postmerger frequency $f_\mathrm{peak}$ yields information about the EoS regime probed in the massive merger remnant, i.e. it is characteristic of the high-density regime of the EoS. Also this GW feature will be measurable with sufficient precision in the future. A significant increase of $f_\mathrm{peak}$ with respect to an empirical relation between $\Lambda$ and $f_\mathrm{peak}$ which holds for purely hadronic EoSs, is indicative of a phase transition.

A strong first-order phase transition after the merger suddenly softens the EoS. This leads to a strong compactification of the remnant, which cannot result from any hadronic model since those do not admit such a strong softening of the EoS. A more compact remnant implies a higher oscillation frequency and thus GW frequency. Loosely spoken, the inspiral does not know yet about the later occurrence of quark matter and the sudden softening of the EoS at higher densities. If a phase transition takes place during merging as the densities increase, the postmerger frequency is then much higher as expected based on the tidal deformability probing the EoS during the inspiral. 

Along these lines we have extended our previous studies here by considering hybrid mergers, where the increase of the postmerger frequency is at least not obvious based on these arguments. Nevertheless, we find that also hybrid merger, i.e. systems where quark matter is already present during the inspiral, lead to a characteristic increase of the postmerger GW frequency. Hence, also within such a scenario the comparison between $\Lambda$ and $f_\mathrm{peak}$ indicates the onset of quark deconfinement. This demonstrates the extraordinary scientific value of GW instruments with dedicated capabilities to detect postmerger frequencies in the range of a few kHz~\cite{Hild2011,Martynov2019}.

We suspect that for hybrid mergers mostly temperature effects are responsible for the frequency increase of the postmerger phase. However, we remark that hybrid mergers should be investigated within a more extensive, systematic study to corroborate this signature also for other hybrid EoS models. Note that this concerns mostly relatively extreme models with a very early onset with $M_\mathrm{onset}$ below $\sim 1.1~M_\odot$. For higher $M_\mathrm{onset}$ we expect that there will be merger configurations leading to a very robust signature resulting from a ``hadronic'' inspiral and an ``unexpectedly'' compact remnant with quark core and associated GW frequency increase as discussed in~\cite{Bauswein2019}. Phase transitions with low onset density and with $M_\mathrm{onset}$ somewhat larger than $\sim 1.1~M_\odot$ may in particular be detectable by the very different inspiral behavior of binaries with stellar masses above and below $M_\mathrm{onset}$~\cite{Chen2019,Chatziioannou2019}. 

We also mention that the mass ejection of mergers undergoing a phase transition may not be too different from purely hadronic mergers~\cite{Bauswein2019b} indicating that the electromagnetic counterpart may not differ strongly~\cite{Metzger2019}. However, the detection of electromagnetic counterparts may yield information about the collapse behavior and thus indirectly about the occurrence of a phase transition~\cite{Bauswein2020}.

%
%
\begin{acknowledgement}
Acknowledgments: We thank N.-U. Bastian for providing EoS tables. We thank T. Fischer and D. Blaschke for helpful discussions. This work was funded by Deutsche Forschungsgemeinschaft (DFG, German Research Foundation) - Project-ID 279384907 - SFB 1245 and - Project-ID 138713538 - SFB 881 (``The Milky Way System'', subproject A10). AB acknowledges support by the European Research Council (ERC) under the European Union's Horizon 2020 research and innovation programme under grant agreement No. 759253. 
\end{acknowledgement}

\end{document}